\documentclass[aps,prd,reprint,groupedaddress]{revtex4-1}
\pdfoutput=1

\usepackage{amsmath}
\usepackage[utf8]{inputenc}
\usepackage{multirow}
\usepackage{graphicx}

\usepackage[colorlinks]{hyperref}
\hypersetup{allcolors=[rgb]{0,0,1}}
\usepackage[dvipsnames]{xcolor}

\begin{document}

\title{Identifying rotation in SASI-dominated core-collapse supernovae with a neutrino gyroscope}

\author{Laurie Walk}
\affiliation{Niels Bohr International Academy and DARK, Niels Bohr Institute, University of Copenhagen, Blegdamsvej 17, 2100, Copenhagen, Denmark}
\author{Irene Tamborra}
\affiliation{Niels Bohr International Academy and DARK, Niels Bohr Institute, University of Copenhagen, Blegdamsvej 17, 2100, Copenhagen, Denmark}
\author{Hans-Thomas Janka}
\affiliation{Max-Planck-Institut f\"ur Astrophysik,
 Karl-Schwarzschild-Str.~1, 85748 Garching, Germany}
 \author{Alexander Summa}
\affiliation{Max-Planck-Institut f\"ur Astrophysik,
 Karl-Schwarzschild-Str.~1, 85748 Garching, Germany}

\date{\today}
\begin{abstract}
Measuring the rotation of core-collapse supernovae (SN) and of their progenitor stars is extremely 
challenging. Here it is demonstrated that neutrinos may potentially be employed as stellar 
gyroscopes, if phases of activity by the standing accretion-shock instability (SASI) affect the neutrino 
emission prior to the onset of the SN explosion. This is shown by comparing the neutrino emission properties of self-consistent,
three-dimensional (3D) SN simulations of a $15\,M_\odot$ progenitor without rotation as well
as slow and fast rotation compatible with observational constraints. The explosion of the
fast rotating model gives rise to long-lasting, massive polar accretion downflows with
stochastic time-variability, detectable e.g.\ by  the 
IceCube Neutrino Observatory for any observer direction. While spectrograms of the neutrino event rate of non-rotating SNe feature a well-known sharp peak due to SASI for observers located in the proximity of the SASI plane, the corresponding spectrograms of rotating models show activity over a wide range of frequencies, most notably above 200 Hz for rapid rotation. In addition, the Fourier power spectra of the event rate for  rotating models exhibit a SASI peak with lower power than in non-rotating models. The spectra for the rotating models also show secondary peaks at higher frequencies with greater relative heights compared to the main SASI peak  than for 
non-rotating cases. These rotational imprints will be detectable for SNe at 10 kpc or closer. 
\end{abstract}

\maketitle

\section{Introduction} 
The influence of angular momentum on the mechanism of core-collapse supernovae (SNe) is a subject of vivid debate \cite{Janka:2017vcp,Janka:2012wk}. Calculations following the evolution of rotating massive stars from the zero-age main sequence until the onset of iron-core collapse showed that rotation changes the stellar structure and is tightly coupled to the magnetic-field evolution~\cite{Heger:2004qp}. The presence of significant angular momentum during stellar core collapse can amplify the magnetic field and can thus influence the SN explosion mechanism, the pulsar spin period~\cite{Ott:2005wh}, and the nucleosynthesis of heavy elements~\cite{Winteler:2012hu}. However, asteroseismological measurements of the interior rotation of low-mass stars suggest core rotation that is slower than predicted by stellar evolution theory~\cite{Cantiello:2014uja}. 

With the recent advent of modern 3D SN modeling with energy-dependent neutrino transport~\cite{Janka:2016fox}, first self-consistent simulations, partly with successful explosions, have become available \cite{Hanke2013,Tamborra:2014hga,Janka:2016fox,Summa:2017wxq,Melson2015a,Melson:2015spa,Mueller2015a,Muller:2016izw,Chan:2017tdg,Takiwaki2014a,Takiwaki:2016qgc,Muller:2017hht,Lentz:2015nxa,Ott:2017kxl}. Besides progenitor perturbations \cite{Couch:2013coa,Muller:2017hht} and previously missing microphysics \cite{Melson:2015spa,Bollig:2017lki}, rotation was found to facilitate explosions by spiral-SASI and triaxial instabilities \cite{Nakamura:2014fxa,Takiwaki:2016qgc,Summa:2017wxq}. Rotation is a key requisite also for the magnetorotational mechanism, where magnetic fields are amplified to dynamically relevant strengths due to extremely high spin rates \cite{Mosta:2014jaa,Nishimura2015,Obergaulinger2017,Obergaulinger2018}. This, however, is at odds with our present theoretical expectation for the far majority of ordinary SNe \cite{Heger:2004qp,Ott:2005wh} and with asteroseismological constraints \cite{Cantiello:2014uja,Ott:2005wh}.

Neutrinos and gravitational waves are  produced deep in the SN core \cite{Janka:2012wk,Janka:2017vcp,Mirizzi:2015eza} and can thus be unique heralds of the core rotation. Neutrinos are the key ingredient in the delayed neutrino-driven explosion mechanism~\cite{Bethe:1984ux}, and their detection can provide crucial insights on the blast-wave dynamics. 
Just before the explosion, during the accretion phase,
the neutrino signal carries characteristic quasi-periodic time modulations \cite{Lund:2010kh,Lund:2012vm,Tamborra:2013laa,Tamborra:2014hga,Kuroda:2017trn,Mueller:2014rna} caused by the standing accretion shock instability (SASI) \cite{Blondin:2002sm, Scheck:2007gw}, replicating the sloshing of the stalled shock wave as it acquires energy.  The first self-consistent SN simulations of non-rotating progenitors in 3D, suggested that the Fourier power spectrum of the neutrino event rate expected from a Galactic burst will show a sharp peak corresponding to the SASI frequency \cite{Tamborra:2013laa,Tamborra:2014hga}. Conversely, when the pre-explosion dynamics is dominated by convection~\cite{Bethe:1990mw}, only high-frequency modulations will appear~\cite{Tamborra:2013laa,Tamborra:2014hga}. 

Grasping the effects of SN rotation on neutrinos is also important because rotation may affect the flavor conversions in the SN envelope~\cite{Mirizzi:2015eza}. Also, the self-sustained lepton asymmetry (LESA), a hydrodynamical instability  fostered by neutrinos~\cite{Tamborra:2014aua}, is damped by progenitor rotation~\cite{Janka:2016fox} with yet unclear consequences. 
Previous work on neutrino signatures of the progenitor rotation was limited by octant symmetry and approximated neutrino transport~\cite{Ott:2012kr}. Recently, a neutrino light-house effect was found due to the T/$|\mathrm{W}|$ instability facilitated by extremely rapid core rotation, much faster than considered here~\cite{Takiwaki:2017tpe}. 

In this paper, we show that SN neutrinos  might serve as stellar gyroscopes by carrying characteristic imprints of progenitor rotation. 
We compare the neutrino  properties of three self-consistent 3D hydrodynamical simulations of a $15\,M_\odot$ SN published by~\cite{Summa:2017wxq}; a non-rotating model, and two models rotating at different velocities compatible with observations.
Distinctive features due to rotation are observable in neutrinos for  Galactic SNe, especially if the observer is located in the proximity of the SASI plane.

\section{Dynamics of rotating supernova models}\label{sec:simulations}
The  hydrodynamical simulations of the $15\,M_\odot$ SN progenitor~\cite{Summa:2017wxq} were carried out on an axis-free Yin-Yang grid~\cite{2004GGG.....5.9005K} with 2~degrees (4~degrees) angular resolution for the rotating (non-rotating) models. 
For all models, the \textsc{Prometheus-Vertex} code with a three-flavor, energy dependent and ray-by-ray-plus neutrino transport with state-of-the-art modeling of the microphysics was used. The nuclear equation of state of Lattimer and Swesty~\cite{Lattimer:1991nc} was applied with a nuclear incompressibility of 220\,MeV. 

The ``non-rotating model'' shows a dynamical behavior similar to other 3D simulations of non-rotating SNe~\cite{Hanke:2013jat,Tamborra:2014aua,Tamborra:2014hga}. A strong SASI spiral mode kicks in at $\sim$150\,ms post-bounce and lasts until 250\,ms to decay in response to the shock expansion after the infall of the Si/SiO interface. SASI motions are largely confined to a well defined plane. Correspondingly, the neutrino emission properties show large-amplitude quasi-sinusoidal time modulations in the proximity of the SASI plane and reduced modulations otherwise~\cite{Tamborra:2013laa}.   

The ``slowly rotating model'' uses the initial angular velocity profile obtained in stellar evolution calculations~\cite{Heger:2004qp}, which include angular momentum transport by magnetic fields. It has an angular momentum of $\sim$$6\times 10^{13}$\,cm$^2$\,s$^{-1}$ at the Si/Si-O interface and a spin period of 6000\,s, leading to a neutron star (NS) with spin period of $\sim$11\,ms, if angular momentum is conserved during collapse. 
This model exhibits a smaller shock radius than the non-rotating one~\cite{Summa:2017wxq}, but a stronger interplay between SASI and convection. This is reflected in the neutrino properties that are comparable to the ones of the non-rotating model in magnitude, but display broadened SASI modulations of smaller amplitude superposed on modulations of higher frequency due to convection.

The ``fast rotating model'' is the only model which successfully explodes. It relies on a pre-collapse rotation profile from~\cite{Mueller:2003fs,Buras:2005tb,Marek:2007gr} with a specific angular momentum of $\sim$$2\times 10^{16}$\,cm$^2$\,s$^{-1}$ at the Si/Si-O interface, i.e., a spin period of 20\,s (corresponding to $\sim$1--2\,ms of the NS for conserved angular momentum).
Notably, the fast rotation favors the growth of an early and strong SASI spiral mode~\cite{Yamasaki:2007dc,2007Natur.445...58B,2012ApJ...749..142F} in a plane perpendicular to the rotation axis, triggering a successful explosion at $\sim$200\,ms  post bounce when the infalling Si/Si-O interface reaches the shock wave~\cite{Summa:2017wxq}. SASI motions drive the shock to large radii in the proximity of the SASI plane leading to an oblate deformation. Because the shock is pushed outwards, the volume of the gain layer increases, thus permitting  higher neutrino heating and facilitating the explosion. 
The increased rotational speed is responsible for neutrino luminosities and mean energies lower than in the other two SN cases because of a reduced mass flow to the proto-NS and lower neutrinospheric temperatures~\cite{Summa:2017wxq}.

The post-explosion dynamics of the fast rotating model are characterized by long-lasting, unsteady 
massive polar accretion downflows, whose corresponding neutrino emission dominates at the
poles. Since matter flows around the NS while radiating neutrinos, the neutrino modulations
 by  polar accretion are visible from all directions.

The neutrino emission properties were extracted at 500\,km and mapped from the Yin-Yang grid onto a regular spherical grid. They were projected to how they would be seen by distant observers located at chosen angular coordinates~\cite{Tamborra:2014aua,Muller:2011yi}. The neutrino energy distributions were then assumed non-thermal~\cite{Keil:2002in, Tamborra:2012ac}.


\section{Detectable modulations of the neutrino signal}\label{sec:detection}
The neutrino telescope  currently providing the largest event statistics is the IceCube Neutrino Observatory~\cite{Abbasi:2011ss}. The main neutrino detection channel is inverse beta decay ($\bar\nu_e+p\to n+e^+$)~\footnote{The detection occurs through Cherenkov radiation of the positron. In rare cases,  multiple photons may be collected for a single neutrino event~\cite{Demiroers:2011am}. However, this scenario will not be considered here.}. IceCube has 5160 optical modules; each module has a background rate of $286$~Hz including dead time effects. Hence, the total background rate is  $R_{\mathrm{bkgd}} \simeq 1.48 \times 10^3$~ms$^{-1}$.  Notably, the $1\sigma$ random fluctuations on the signal will be on average $\sqrt{1.48\times10^3}\,\mathrm{ms}^{-1}=38.5$~ms$^{-1}$, which is  lower than neutrino-signal modulations of $\mathcal{O}(100\ \mathrm{ms}^{-1})$ for a Galactic SN. Modulations of the neutrino signal due to the SN pre-explosion dynamics will therefore be detectable.
The neutrino signal observable by IceCube has been estimated following~\cite{Abbasi:2011ss,Tamborra:2014aua}, with an effective correction term which takes into account detection channels other than inverse beta decay, and using the inverse beta decay cross section as presented in~\cite{Strumia:2003zx}. For more details, we refer the interested reader to Refs.~\cite{Tamborra:2013laa,Tamborra:2014aua,Abbasi:2011ss}.

Neutrinos change their flavor as they propagate in the SN envelope~\cite{Mirizzi:2015eza}. The exact flavor 
composition reaching the observer is unclear, especially given the most recent developments, see e.g.~\cite{Tamborra:2017ubu,Izaguirre:2016gsx,Sawyer:2015dsa,Sawyer:2005jk}. In view of these uncertainties, 
 we will neglect flavor conversions and  rely on two extreme scenarios; one where a complete flavor conversion of $\bar\nu_e$ in $\bar\nu_x$ (with $x$ being $\mu$ or $\tau$) occurs, and one case where $\bar\nu_e$ do not oscillate and are left unchanged. Any other scenario will fall between the two extremes. 

\begin{figure}
\centering
\includegraphics[width=\columnwidth]{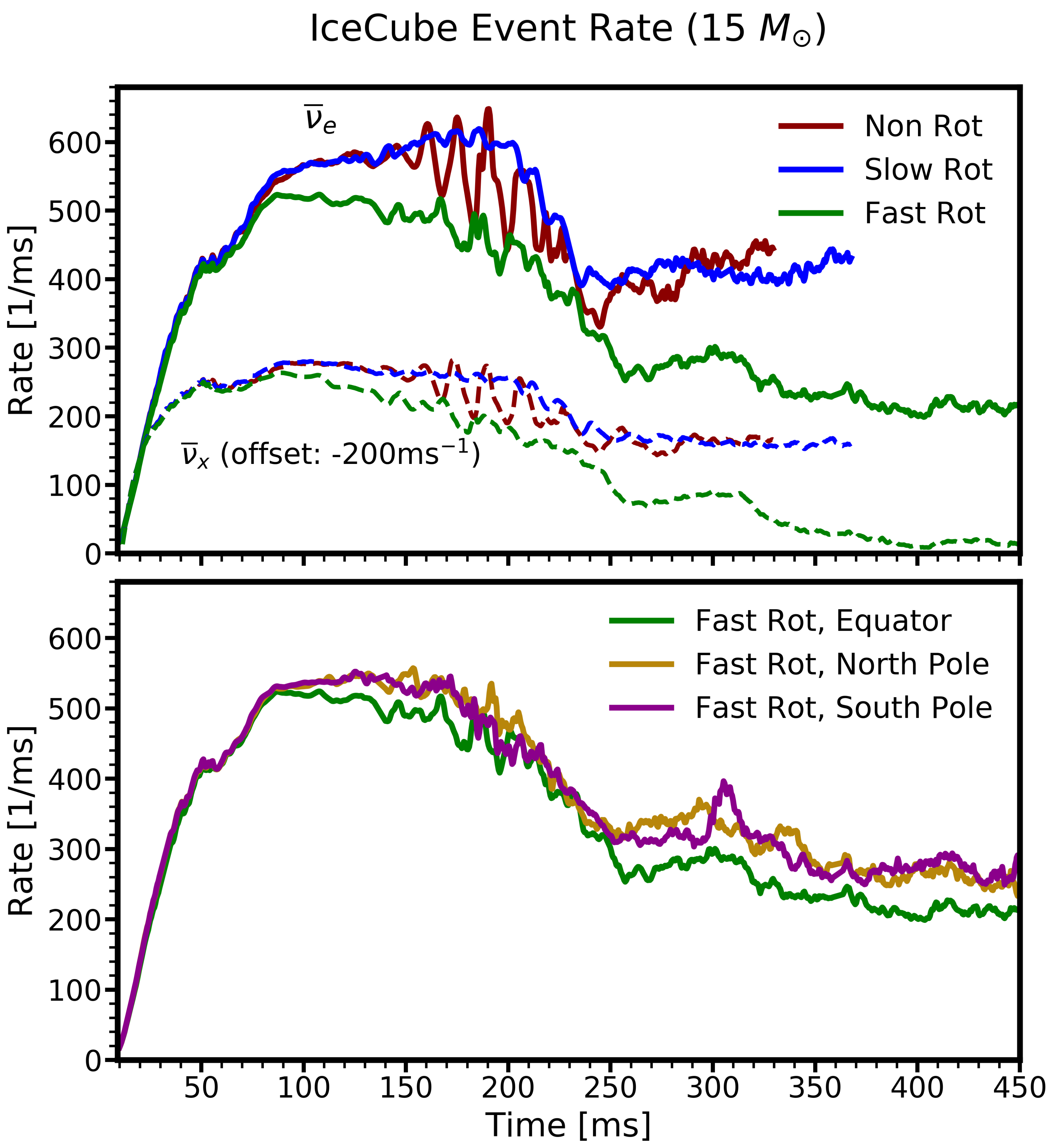}
\caption{IceCube detection rate for the three $15\,M_\odot$ SN cases placed at 10\,kpc from the observer. {\it Upper panel}: Event rate for the non-rotating, slowly and fast rotating progenitors. The observer direction is chosen in favor of strong signal modulations. Two scenarios for flavor conversions ($\bar\nu_e$ for no flavor conversion and  $\bar\nu_x$
for complete flavor conversion) are shown.  {\it Bottom panel}: $\bar\nu_e$ event rate for the fast rotating SN for an observer located along the SN rotation axis and on a plane perpendicular to it.}
\label{fig:rates}
\end{figure}

The event rates for the three investigated $15\,M_\odot$ SN progenitors are displayed in the top panel of Fig.~\ref{fig:rates}. The event rate for the non-rotating SN model, for a distant observer located in the proximity of the SASI plane, exhibits clear modulations due to the SASI spiral motions, similar to what was found for 27 and 20\,$M_\odot$ models in~\cite{Tamborra:2014aua,Tamborra:2013laa}. Both scenarios of full (or absence of) flavor conversion reveal
 quasi-periodic modulations of the signal. The neutrino event rate of the slowly rotating SN model is on average comparable to the one of the non-rotating progenitor, given that both models do not explode. It still shows   modulations due to SASI, but these are smeared and weakened by rotation and convection. 

\begin{figure}
\centering
\includegraphics[width=\columnwidth]{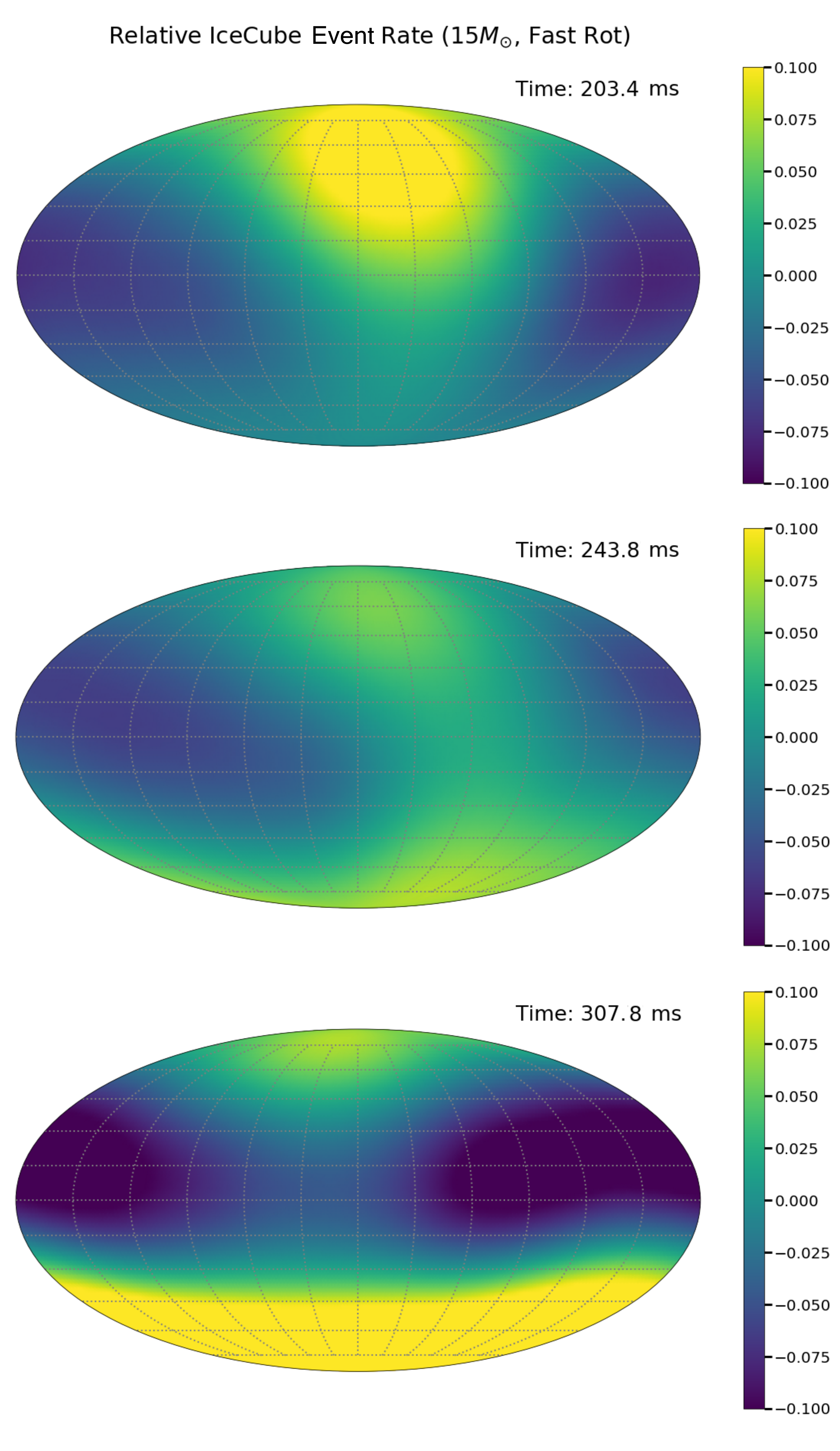}
\caption{Directional variations of the normalized $\bar\nu_e$ IceCube rate, $[(R-\langle R\rangle)/\langle R\rangle]$, for the fast rotating $15\,M_\odot$ SN progenitor on a map of observer directions at three post-bounce times. 
}
\label{fig:fastmaps}
\end{figure}

In the rapidly rotating progenitor, the average event rate is lower than in the other two cases because of the lower energy budget radiated in neutrinos due to the quenched accretion by the onset of the explosion.
The neutrino signal clearly reflects the unsteady downflow dynamics and does not exhibit any clean SASI modulation. To better illustrate these characteristic features, Fig.~\ref{fig:fastmaps} shows, on a Mollweide map of observer directions, snapshots at different times of the amplitude of the $\bar\nu_e$ IceCube event rate $R$, normalized  by its $4\pi$ average $\langle R\rangle$: $[(R-\langle R\rangle)/\langle R\rangle]$. The downflows are visible by the relative event rate along the rotation axis being higher than in the perpendicular plane. This is further illustrated in the bottom panel of Fig.~\ref{fig:rates}, where the neutrino event rates seen along the polar directions and in the perpendicular plane exhibit similar modulation patterns, and the overall event rate is higher along the rotation axis.  
\href{https://wwwmpa.mpa-garching.mpg.de/ccsnarchive/data/Walk2018/}{Movies} of the time-evolution of
the event rate for the three studied cases are also provided as supplementary material.

We focus on IceCube for its large event statistics. However, Hyper-Kamiokande~\cite{Abe:2018uyc} could provide similar information with lower statistics, but guaranteeing a background-free signal that will be competitive for SNe at large distances~\cite{Tamborra:2013laa}. Notably, a combined reconstruction of the neutrino light-curve observed by IceCube, Hyper-Kamiokande, and possibly DUNE~\cite{Acciarri:2015uup} would be optimal to better pinpoint the features described above.

It is worth noticing that Hyper-Kamiokande will provide energy information, in addition to details on the temporal evolution of the neutrino signal. However, in the following, we will focus on the analysis of the neutrino event rate only. In fact, we found that although the neutrino energy spectra may give us hints on the progenitor rotation as they are more pinched for the fast rotating progenitor than for the slow and non-rotating $15\,M_\odot$, such a trend is degenerate with the progenitor mass~\cite{Walk2018b}. The neutrino energy spectra therefore cannot  help us in discriminating the SN progenitor rotation. 


\section{Fourier analysis of neutrino event rate}
In this Section, after  pinpointing the directions along which the observer should be located to detect the largest modulations of the neutrino signal, we present  a Fourier decomposition of  the neutrino event rate detectable in IceCube. We employ  spectrograms of the neutrino signal for the rotating and non-rotating progenitors and  highlight the  features characteristic of rotation. We then investigate the Fourier power spectrum in  the  time  window with the largest modulations of the neutrino signal, and further identify detectable traits of rotation.

\subsection{Time evolution of the neutrino event rate}
In order to pinpoint the directions of strongest and weakest variation of the neutrino event rate, we introduce for each angular location of the observer  the time-dependent standard deviation of the
IceCube event rate relative to  the average of the expected rate over all
possible observer directions ($\langle R \rangle$): $\sigma^2 = \int_{t_1}^{t_2} dt [(R-\langle R \rangle)/\langle R \rangle]^2$~\cite{Tamborra:2014hga}.
\begin{figure*}
\centering
\includegraphics[width=0.8\textwidth]{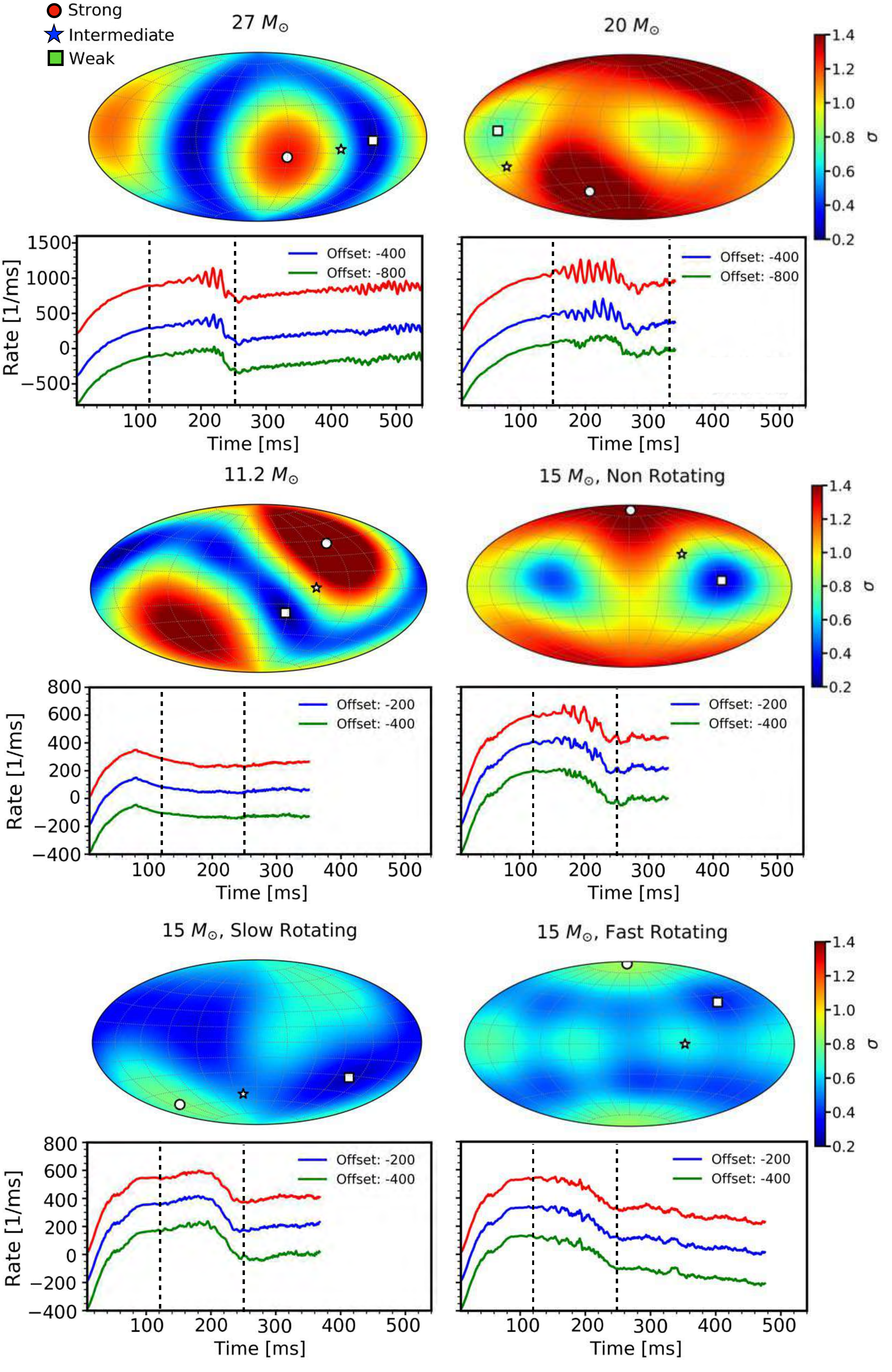}
\caption{Molleweide maps of the standard deviation of the IceCube $\bar\nu_e$ event rate, $\sigma$, integrated over the time interval [120, 250]~ms for the $27, 11.2$, and the three $15\,M_\odot$ SN models, and over [150, 330] ms for the $20\,M_\odot$ SN model. On each map, three directions where strong, intermediate and weak modulations of the neutrino signal occur, are marked with a  circle, a  star and a square, respectively.  The corresponding event rates for a SN at 10 kpc, along each of  the three selected directions, are shown underneath each map without detector shot noise. The red, blue, and green curves correspond to the event rates along the ``strong",  ``intermediate", and ``weak" directions, respectively. Vertical dotted lines indicate the boundaries of the time intervals integrated over for each of the models. The rotating models show weaker signal modulations than the non-rotating models. The fast-rotating model exhibits a quadrupolar pattern along the equatorial plane due to rotation.}
\label{fig:maps}
\end{figure*}

Figure~\ref{fig:maps} shows Molleweide maps of $\sigma$ for the three  $15\,M_\odot$ progenitors first presented in this work. The regions with the largest (smallest) rate variations of the neutrino signal relative to the average rate are the ones displaying the hottest (coldest) regions on the maps, corresponding to the red-yellow (green-blue) colors. 
  In order to facilitate a comparison between rotating and non-rotating models, we also include the  three non-rotating
progenitors investigated in Refs.~\cite{Tamborra:2014aua,Tamborra:2013laa,Tamborra:2014hga} with masses of $11.2, 20$, and $27\,M_\odot$.  The 
$27\,M_\odot$ progenitor has two SASI episodes (in [120, 250]~ms and [410, 540] ms) with a convection-dominated phase in between, while the $20\ M_\odot$ model  has only one strong SASI episode (in [150, 330]~ms), and the  $11.2\,M_\odot$ is clearly  convection-dominated~\cite{Tamborra:2014aua,Tamborra:2013laa,Tamborra:2014hga}.
As discussed in Ref.~\cite{Tamborra:2014hga} and visible from Fig.~\ref{fig:maps}, the SASI modulations of the neutrino signal are stronger in the proximity of the SASI plane. 
Among all studied progenitors,  the $20\,M_\odot$ SN model has the strongest SASI activity. A quadrupolar pattern of the standard deviation is caused by the dipolar asymmetry of the $\bar{\nu}_e$ emission associated with the LESA instability~\cite{Tamborra:2014aua}  in the $11.2\,M_\odot$ SN progenitor;  the LESA dipole direction is located in the upper right  red region in the map.

The two rotating progenitors (shown in the bottom of Fig.~\ref{fig:maps}) exhibit the smallest $\sigma$ variations, indicating  weaker SASI modulations in rotating models compared to the non-rotating ones.  The maxima in the map of the  slow rotating model correspond to the directions with the largest long-time variability of the signal, and as will be evident from Fig.~\ref{fig:powerspectrum}, these directions do not  coincide with the ones with the strongest  SASI modulations. This is because the SASI modulations in this model are relatively weak, and $\sigma$ is dominated by other time-dependent variations.  Interestingly, the fast rotating model shows a clear quadrupolar structure along the equatorial plane, perpendicular to the rotation axis. This is a consequence of spiral-SASI modulations of accretion flow in the equatorial plane. The unsteady polar downflows discussed in Sec.~\ref{sec:detection} are  responsible for the hottest regions at the poles (see also Fig.~\ref{fig:fastmaps}). 

The neutrino event rates for a SN at 10 kpc from the observer are shown underneath each Molleweide map in Fig.~\ref{fig:maps}, in order to better illustrate the differences in modulations seen by observers located at different angular directions. The selected directions are marked  on the maps with a  circle, a  star, and a  square, corresponding to the directions of  ``strong", ``intermediate", and ``weak" modulations of the signal, respectively. The ``strong'' (``weak'') directions have been chosen by locating the maximum (minimum) of $\sigma$ on each map.

\subsection{Spectrograms of the neutrino event rate}

In order to investigate whether rotation induces modulations in the neutrino signal throughout the duration of the SN simulation, we compute the Short Time Fourier Transform of the neutrino event rate. We  follow Ref.~\cite{Lund:2010kh} with a Hann window function minimizing the edge effects. 
The Fourier decomposition is computed in a running window of width $\tau= 50$\,ms, which slides through the signal in steps of 1\,ms. The spacing of the discrete Fourier frequencies is $\delta f=1/\tau=20~{\rm Hz}$.

\begin{figure*}[t!]
\centering
\includegraphics[width=\textwidth]{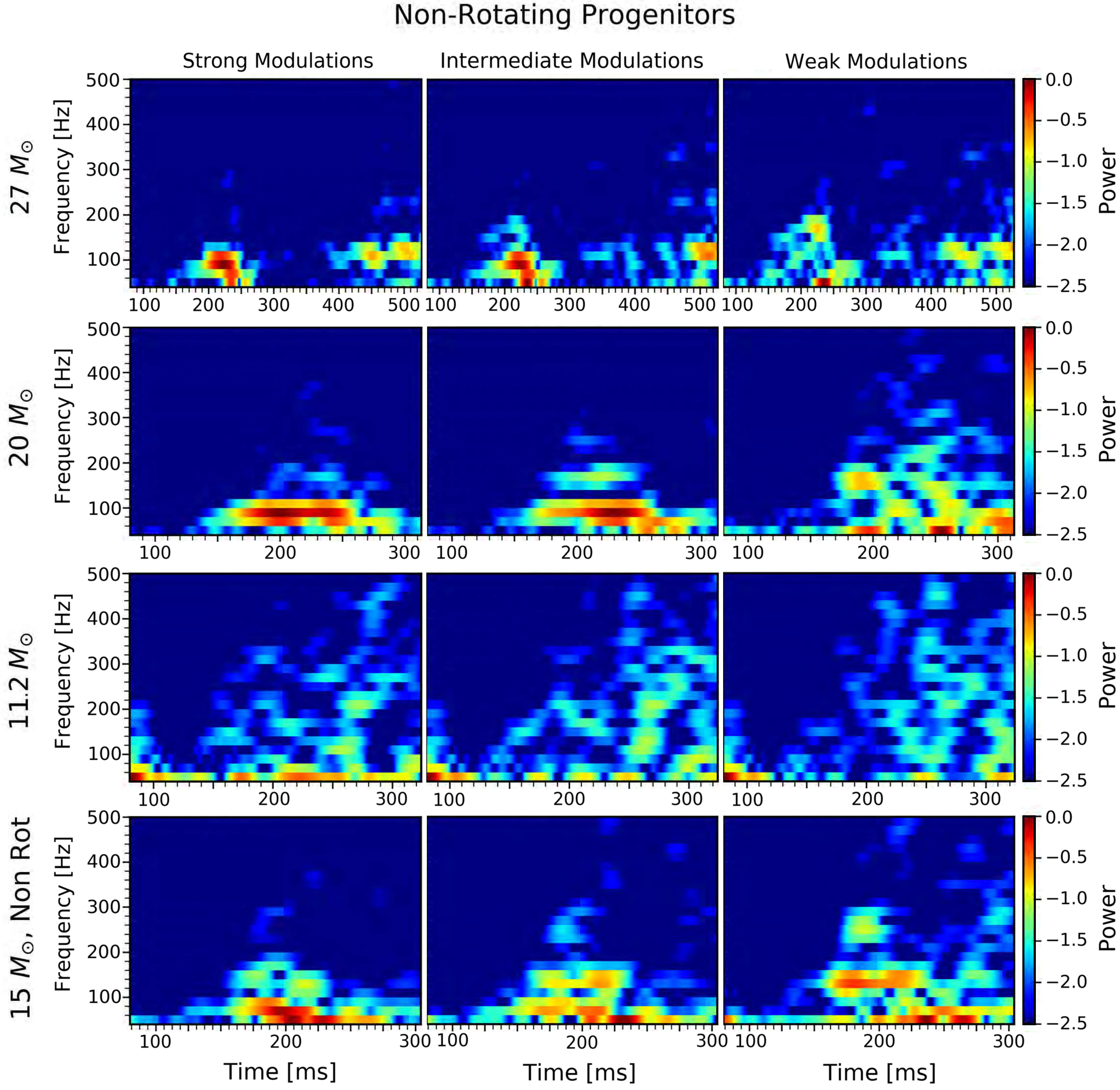}\\
\caption{Spectrograms of the $\bar\nu_e$ event rate  observed in IceCube along the directions selected in Fig.~\ref{fig:maps} with strong, intermediate and weak modulations of the signal (from left to right)  for all non-rotating models (i.e., $27, 20, 11.2$, and 
$15\,M_\odot$ from top to bottom). The spectrogram power has been normalized to the maximum Fourier power along the selected observer direction, and plotted on a log color scale.  The SASI activity corresponds to the hottest regions in each spectrogram for the $27, 20$ and 
$15\,M_\odot$  models. Convection is instead characterized by signal variations more uniformly  spread over all frequencies above 50~Hz in the $11.2\,M_\odot$ model. }
\label{fig:spectrogr_nonrot}
\end{figure*}
%
\begin{figure*}[t!]
\centering
\includegraphics[width=\textwidth]{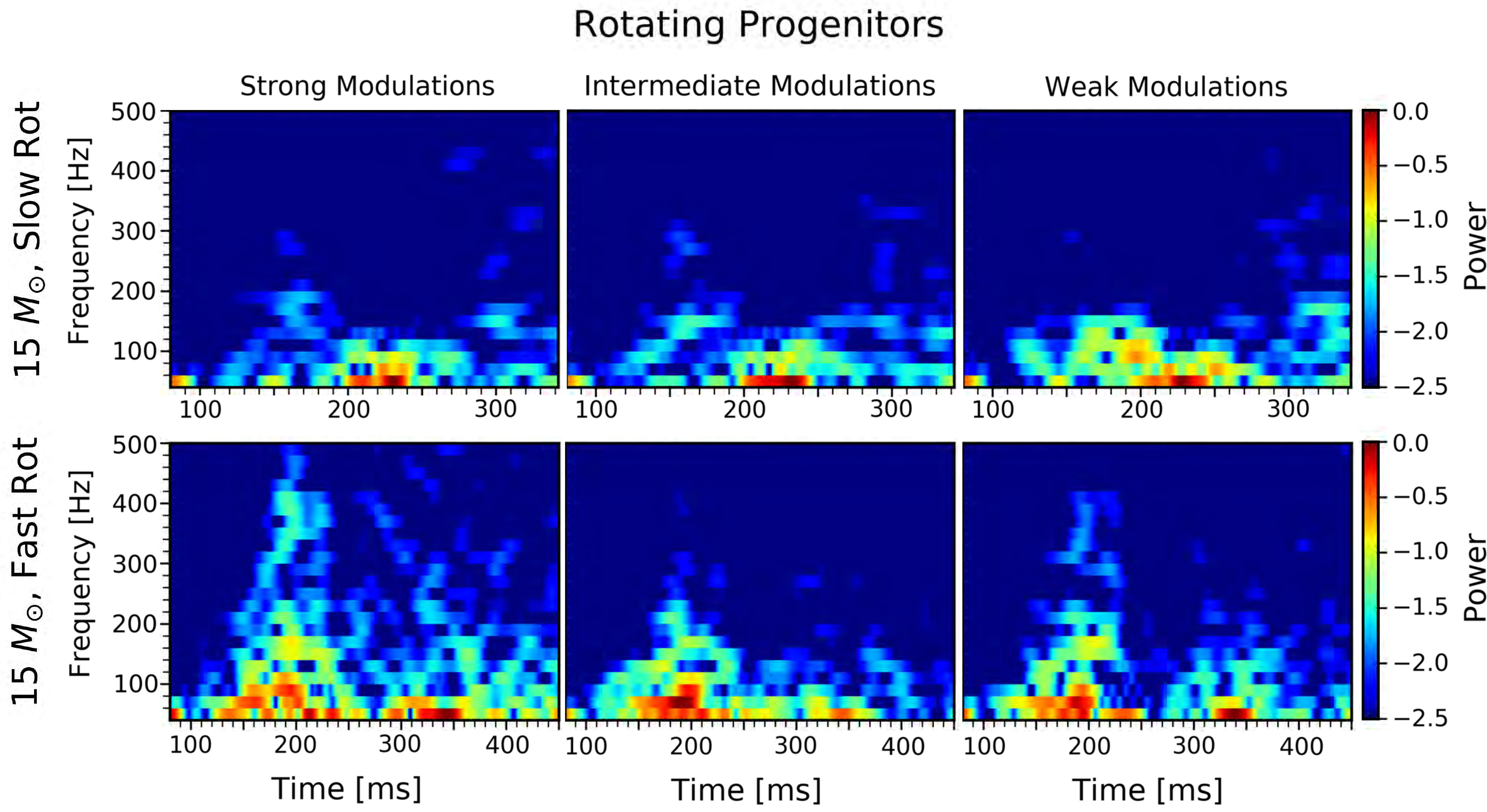}
\caption{Analogous to Fig.~\ref{fig:spectrogr_nonrot}, but for the two rotating progenitors (i.e, $15\,M_\odot$ slowly rotating on the top and  $15\,M_\odot$ fast rotating on the bottom). Because of rotation, the spectrograms  for the ``strong modulation'' directions show activity above 200~Hz and not limited to time intervals where SASI occurs.}
\label{fig:spectrogr_rot}
\end{figure*}
Figure~\ref{fig:spectrogr_nonrot} shows the resulting spectrograms of the $\bar\nu_e$ IceCube event rate for the non-rotating models in the three observer directions selected in Fig.~\ref{fig:maps}. The directions with strong, intermediate and weak signal modulations are shown from left to right respectively. Each spectrogram has been normalized to the maximum Fourier power along that selected observer direction for the considered SN model. 

Due to the SASI activity,  the spectrograms of the $27, 20$, and 
$15\,M_\odot$ models show  hot regions around $75-100$~Hz during the time interval  of the SASI. Interestingly, even in the direction with the weakest signal modulation, signatures of SASI  can be identified as modulations around $150-200$~Hz (i.e., at frequencies about two times higher than the typical SASI frequency) in the SASI time window. This feature is not directly visible by looking at the neutrino event rate, yet it is highlighted in the spectrograms. As the SASI activity is less dominant in the direction of weak modulation, the spectrograms in this direction also exhibit a wider spread in other frequencies. 

As for the intermediate directions, the hot  SASI regions  are still visible, although less pronounced than in the spectrograms corresponding to strong signal modulation. None of the spectrograms displays significant signal modulations above 300~Hz. The $11.2\,M_\odot$ SN model, in contrast to the other non-rotating models, is characterized fully by convection, which is visible in the spectrogram as low-amplitude power more uniformly distributed over time through all frequencies up to $400-500$~Hz.

For comparison, Fig.~\ref{fig:spectrogr_rot} shows the spectrograms for the slowly and fast rotating models along the three selected observer directions. In contrast to the non-rotating progenitors, the spectrograms of the rotating progenitors in the strong-modulation direction show a wide spread in frequencies other than the SASI frequency (see Figs.~\ref{fig:rates} and Fig.~\ref{fig:maps}). This may be an illustration of a more intricate interplay between SASI and convection due to rotation, as will be further discussed in the next section. 
Along the ``strong" direction, the spectrograms show significant activity  above 200 Hz, not present in the corresponding direction for the non-rotating progenitors. Although this feature signals rotation, it might be hard to identify it unambiguously because of  degeneracies with the signatures seen by observers located away from the SASI plane for  non-rotating progenitors (see, e.g.,~the spectrogram of the weak direction of the $20\,M_\odot$ in Fig.~\ref{fig:spectrogr_nonrot}). To overcome this challenge in using the neutrino signal to identify rotation, other detectable features must be employed and combined with the spectrograms.

\subsection{Fourier transform of neutrino event rate}
\begin{figure*}[t!]
\centering
\includegraphics[width=\textwidth]{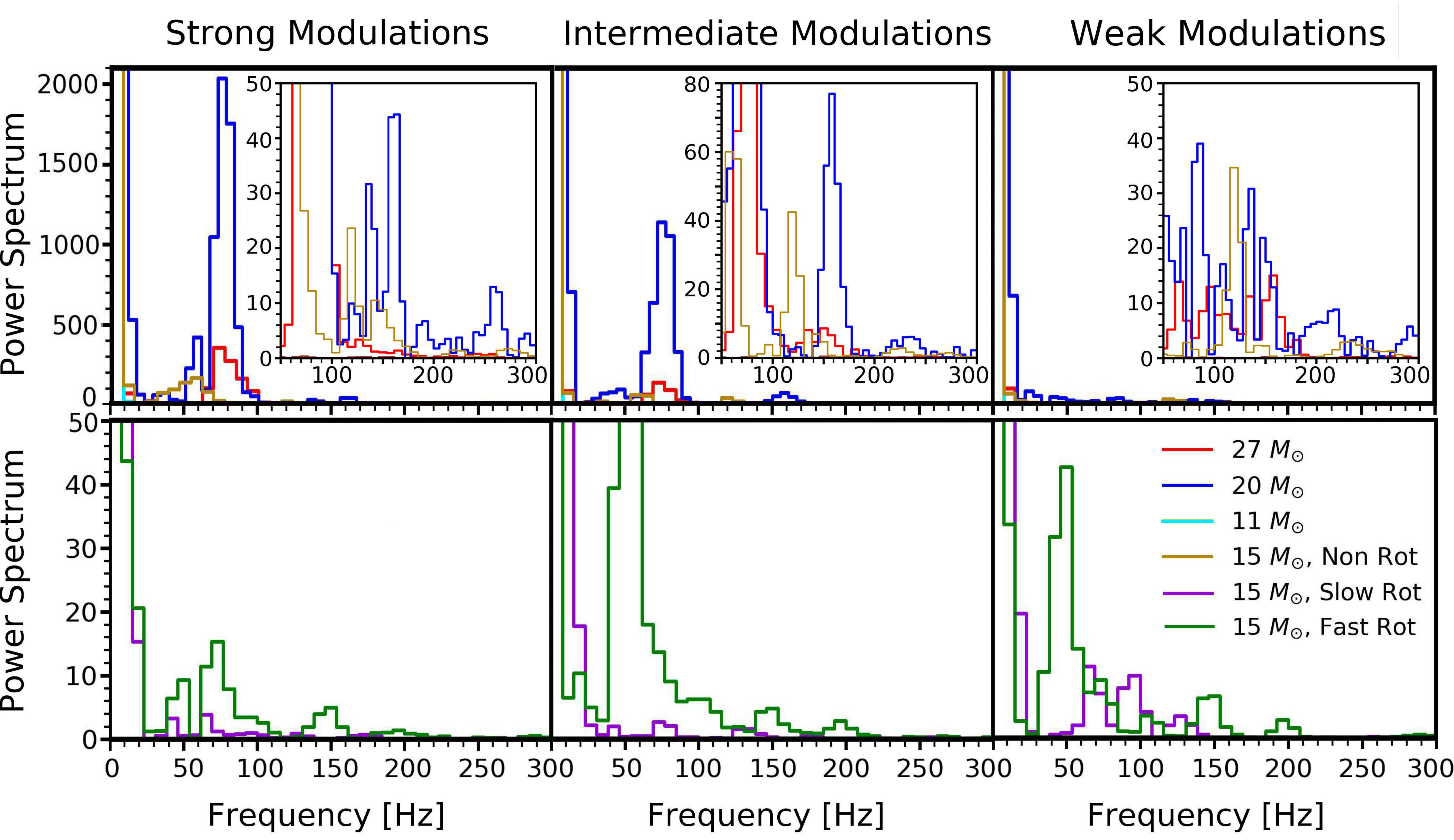}
\caption{Power spectra of the $\bar\nu_e$ event rates observed in IceCube for a SN at 5 kpc, normalized to the average power of the IceCube shot noise. The power spectrum has been computed over $[120, 250]$\,ms for the 27, 11.2, and each of the $15\,M_\odot$ models, and over $[150, 330]$\,ms for the $20\,M_\odot$ model. \textit{Upper panels:} Power spectra for the four non-rotating models ($27, 20, 11.2$, and $15\, M_\odot$ progenitors) along the directions selected in Fig.~\ref{fig:maps}.  In each  panel, the inset shows the zoomed-in power spectrum in order to highlight the peaks appearing at frequencies higher than the SASI frequency. A peak characteristic of the  SASI activity appears at $\sim 75$~Hz for the  $27$ and $20\,M_\odot$ progenitors and at $\sim 55$~Hz for the $15\,M_\odot$ progenitor. The insets show secondary peaks corresponding to frequencies which are multiples of the SASI frequency. The $11.2\,M_\odot$ progenitor features a flat power spectrum due to convection. \textit{Bottom panels:} Fourier spectra for the two rotating models of $15\,M_\odot$ are shown. Both display a SASI peak with a significantly lower power than in the cases of the non-rotating progenitors.}
\label{fig:powerspectrum}
\end{figure*}

To distinguish the time-integrated features, Fig.~\ref{fig:powerspectrum} shows the Fourier power spectra of the $\bar\nu_e$ IceCube event rates for a SN at 5 kpc for all progenitor models along the directions selected in Fig.~\ref{fig:maps}. The power spectrum has been computed over the time interval [120, 250]~ms, corresponding to $\delta f=1/\tau \approx 7.7~{\rm Hz}$, for the 27, 11.2, and each $15\,M_\odot$, and over [150, 330] ms, corresponding to $\delta f=1/\tau \approx 5.5~{\rm Hz}$, for the $20\,M_\odot$ model. 

While the $11.2\,M_\odot$ progenitor has a flat power spectrum as a result of convection and the absence of SASI activity, the other three non-rotating models show a main peak corresponding to the SASI frequency around $\sim 75$~Hz for the $27$ and $20\,M_\odot$ progenitors and  $\sim 55$~Hz for the $15\,M_\odot$ progenitor. The difference in the SASI frequencies is due to differences in the shock radii between the progenitors, being related by $f_{\mathrm{SASI}} \propto R_{\mathrm{shock}}^{-3/2}$~\cite{Scheck:2007gw,Tamborra:2013laa}. 

The power of the SASI peak is progenitor dependent, being highest for the $20\,M_\odot$ case, which, as also visible from Fig.~\ref{fig:maps},  has the neutrino event rate with the largest modulations. One can see from the insets in each of the top panels of Fig.~\ref{fig:powerspectrum}, that secondary peaks appear at higher frequencies, and that these are multiples of the SASI frequency. While the SASI peak will be clearly detectable for a SN up to 20~kpc~\cite{Tamborra:2013laa,Tamborra:2014hga}, these secondary peaks will only stand out over the background noise for closer SNe, e.g.~located at $\sim$5~kpc. As the observer moves away from the SASI plane, the neutrino event rate is less modulated (see Fig.~\ref{fig:maps}) and, correspondingly, the SASI peak shows less power.

The bottom panel of Fig.~\ref{fig:powerspectrum} displays the Fourier spectra for the two $15\,M_\odot$ rotating progenitors. As previously concluded from Figs.~\ref{fig:rates} and \ref{fig:maps}, rotation weakens the SASI modulations in the neutrino event rate. This results in SASI peaks with less power than in the case of the $15\,M_\odot$ non-rotating progenitor.

SASI occurs at a higher frequency in the slowly rotating model ($\sim 70$~Hz) because the shock radius is smaller than in the $15\,M_\odot$ non-rotating progenitor~\cite{Summa:2017wxq,Kazeroni:2017fup}, and at a slightly lower frequency for the fast rotating model ($\sim 50$~Hz), because of a larger shock radius (see also discussion in Sec.~\ref{sec:simulations}), although the modulations of the  neutrino event rates of the $15\,M_\odot$ fast rotating model do not appear to depend strongly on the observer direction when examining the plot in Fig.~\ref{fig:rates}. The SASI-associated, quasi-periodic variations are stronger (and visible as higher SASI peaks in Fig.~\ref{fig:powerspectrum}) in the ``intermediate'' and ``weak'' directions. The reason is the fact that in the fast rotating model the ``strong'' direction is located at the poles, whereas the other two observer directions are
closer to the equator, i.e.~closer to the SASI plane.

The SASI peaks also appear to be broader for the rotating models than for the non-rotating cases.   This might be determined by a stronger interplay between SASI and
convection fostered by rotation~\cite{Foglizzo:privcomm,Yamasaki:2007dc}. We stress that such a possibility is only a conjecture;  a theoretical model predicting the 
nonlinear consequences of  rotation on the coupling between buoyant motions and SASI is still missing and it should be subject of further dedicated work.

The power spectra of the rotating models also exhibit higher-frequency peaks at $\sim140$~Hz, $\sim170$~Hz ($\sim150$~Hz, $\sim200$~Hz) for the slowly (fast) rotating model with relative heights compared to  their respective main SASI peak higher than in the non-rotating models. This is especially visible in the fast rotating progenitor, where the higher power of these high-frequency modulations is associated with the neutrino-emission variations associated with the massive, unsteady polar downflows, which dominate the total radiated power in signal modulations (see Figs.~\ref{fig:spectrogr_rot} and \ref{fig:maps}). The difference in the relative heights between peaks of different frequencies represents another signature of rotation detectable through neutrinos.

\section{Conclusions}
Sophisticated SN simulations in 3D of rotating progenitors show signatures in the neutrino signal due to SASI and its interplay with convection and rotation. For the first time, we present a method to identify the rotation of the SN progenitor using neutrinos as gyroscopes, if the SN dynamics are dominated by SASI. We  employ both the spectrograms of the detectable neutrino event rate and the Fourier power spectrum, and identify rotational imprints in each.

For observers located close to the SASI plane, the spectrograms of the neutrino event rate, e.g.\ detectable in IceCube, show modulations more uniformly spread over frequencies (notably above 200 Hz for rapid rotation), in contrast to the localized, strong, low-frequency modulation due to SASI  found in simulations of non-rotating progenitors. Moreover, the Fourier power spectra of the rotating models show secondary peaks at frequencies higher than the SASI one, whose relative heights compared to  the main SASI peak are larger  than in the non-rotating progenitors. These characteristic features in the Fourier spectra will be  detectable for  SNe at $\sim 5\,$kpc, while the variations in the spectrogram will be detectable even for SN at larger distances, but still within our Galaxy. {However, degeneracies may be observed in the case of fast rotating progenitors and progenitors whose main dynamical activity is convection; in this case, multi-messenger observations may aid the discrimination of the rotational properties of the progenitor.}

We forecast the neutrino event rate for the IceCube Observatory because of the best statistics. Nevertheless, signatures due to the progenitor rotation could be detectable for SNe located throughout our Galaxy by combining the event rates observed in multiple upcoming neutrino detectors, such as Hyper-Kamiokande and possibly DUNE.   

\section*{Acknowledgments}
We are grateful to Tobias Melson for providing access to the data of Ref.~\cite{Summa:2017wxq}, and Thierry Foglizzo, R\'emi Kazeroni, Bernhard M\"uller, and Georg Raffelt for insightful discussions. This project was supported by the Villum Foundation (Project No.~13164), the Knud H\o jgaard Foundation, the Danish National Research Foundation (DNRF91), the European Research Council through grant ERC-AdG No.\
341157-COCO2CASA, the Deutsche Forschungsgemeinschaft through Sonderforschungbereich
SFB~1258 ``Neutrinos and Dark Matter in Astro- and
Particle Physics'' (NDM), and the Excellence Cluster ``Universe''
(EXC~153). The model calculations were performed on SuperMUC at
the Leibniz Supercomputing Centre with resources granted
by the Gauss Centre for Supercomputing (LRZ project ID: pr74de).

\bibliography{SNrotating}

\end{document}